\documentclass[twocolumn,twoside,slac_two]{revtex4}
\usepackage{graphicx}
\usepackage{fancyhdr}
\begin{document}

\title{Multi-waveband Variations of Blazars during Gamma-ray Outbursts}

%

\author{A.~P. Marscher}
\affiliation{Institute for Astrophysical Research, Boston University, 725
Commonwealth Ave., Boston, MA 02215, USA}

\begin{abstract}
Multi-wavelength light curves of bright gamma-ray blazars (e.g., 3C~454.3) reveal strong correlations across wavebands, yet striking dissimilarities in the details. This conundrum can be explained if the variable flux and polarization result from both (1) modulation in the magnetic field and relativistic electron content imparted at the jet input and (2) turbulence in the flow. In the Turbulent Extreme Multi-Zone (TEMZ) model being developed by the author, 
much of the optical and high-energy radiation in a blazar is emitted near the 43 GHz core of the jet as seen in VLBA images, parsecs from the central engine, as indicated by observations of a number of blazars. The model creates simulated light curves through numerical calculations that approximate the behavior of turbulent plasma---modulated by random fluctuations of the jet flow---crossing a cone-shaped standing shock system that compresses the plasma and accelerates electrons to highly relativistic energies. A standing shock oriented transverse to the jet axis (Mach disk) at the vertex of the conical shock
can create a variable nonthermal seed photon field that is highly blueshifted in
the frame of the faster jet plasma, leading to highly luminous, rapidly variable $\gamma$-ray 
emission.

\end{abstract}

\maketitle

\thispagestyle{fancy}


\section{\label{intro}INTRODUCTION}

The variability of blazars from radio to $\gamma$-ray frequencies has long been a challenging
problem that theoretical models have yet to solve. The causes of this futility include
complexity in the data, which in the past have been undersampled in both time and frequency.
The routine all-sky survey mode of the {\it Fermi} Large Area Telescope (LAT), plus the 
concerted effort by many research groups to provide commensurate data at other wavebands,
are combining to provide new data sets that are sufficiently rich to both test existing
scenarios for blazar emission and stimulate the design of new models. The hope is that
finding the correct explanation for the variations will allow us to use the data to
infer the structure of the jet as close to the black hole as possible.

The author is leading a collaboration dedicated to amassing extensive multi-waveband
fluxes, polarization, and parsec-scale images over time. This has led to the
recognition that the current array of models with, at most, a few emission zones
are unable to reproduce some prominent aspects of the observational data. This
work summarizes the relevant characteristics of the data and introduces a new
numerical, many-zone scheme that can potentially explain many aspects of the
observed behavior of blazars.

\section{\label{location}LOCATION OF GAMMA-RAY FLARES}

Successful modeling of blazar emission requires that we identify the location(s) where
the flares occur. In order to do this, the author and his collorators use their
extensive multi-waveband data to determine the timing between when a high-energy flare
occurs and when a superluminal knot passes through the ``core'' on millimeter-wavelength
Very Long Baseline Array (VLBA) images. (The core is an approximately stationary, bright,
compact feature at one end of the jet on sub-milliarcsecond (parsec) scales. At millimeter
wavelengths, it has observed properties that are similar to those expected from a
standing conical shock.\citep{caw06,darc07})
They employed this method in an analysis of BL~Lac in 2005 to
determine that there are multiple sites of X-ray and optical
flares in the jet: upstream of the core and within the core.\citep{mar08}
It appears that flares occur as enhanced plasma (higher magnetic field and particle
density) in the jet flow passes through these regions.
Evidence for this is provided by the quasar PKS~1510$-$089, which exhibited a
series of $\gamma$-ray and optical flares in early 2009 as the optical polarization position
angle $\chi$ rotated systematically by $\sim~720^\circ$.
As this rotation ended, a bright superluminal knot passed through the core at (to
within the uncertainties) the date of an extremely bright optical/$\gamma$-ray flare. 
The last flare therefore took place either in or very close to the core.

Further evidence that the majority of flares at $\gamma$-ray (and usually other) frequencies
originate in or near the millimeter-wave core comes from the coincidence in timing of
millimeter-wave and $\gamma$-ray flares.\citep{lt11} However, the $\gamma$-ray peak
usually occurs during the rise of the mm-wave flux, leaving the possibility that
the $\gamma$-ray event takes place somewhat upstream of the core. The author's group
monitors the structure of the jets of 35 Fermi-detected blazars with the VLBA at 7 mm.
After three years of this program, preliminary analysis finds that 43 $\gamma$-ray flares
were simultaneous within errors with the appearance of a new superluminal knot or
a major outburst in the core at 7 mm. This compares with only 13 cases in which either a
$\gamma$-ray flare occurred without a corresponding mm-wave event, or a mm-wave flare or 
superluminal knot ejection had no $\gamma$-ray counterpart. (Four of these blazars had other
$\gamma$-ray flares that {\emph did} correspond to a mm-wave event.) In addition, five
blazars that were quiescent at $\gamma$-ray energies were also quiescent at 7 mm.

The evidence is therefore clear that most (but not all) of $\gamma$-ray flares occur
near or in the mm-wave core. The challenge is then to create a model within this context
that is capable of reproducing the observed characteristics of the multi-waveband variability
of blazars.

\section{OBSERVED PROPERTIES OF BLAZAR VARIABILITY}

Over the past decade, a number of detailed studies of long-term, well-sampled light curves
of blazars at various regions of the electromagnetic spectrum have uncovered some patterns
that should be incorporated in theoretical models. One is the power spectral density
(PSD) of the flux variations, which in blazars is generally has a red noise
spectrum, meaning a power law with steep slope.\citep{chat08,mch10,abdo10} This suggests that a
random process modulates the emission. For synchrotron and inverse Compton radiation, the
random process must govern fluctuations in the magnetic field and
density of relativistic electrons. Turbulence is a likely cause
of such fluctuations. This hypothesis is supported by the degree of linear polarization,
which indicates the level of disorder of the
magnetic field vectors across the emission region. If we approximate the turbulence in terms
of $N$ discrete cells, each with its own magnetic field orientation, the mean degree of
polarization should be $p\sim p_{\rm max}N^{-1/2}$, where $p_{\rm max} \approx 0.75$ is the
fractional polarization in the presence of a uniform magnetic field.\citep{burn66} A typical
optical polarization is of order 10\%, hence we can surmise that there are $\sim 60$
turbulent cells present at optical wavelengths. The polarization tends to be lower at
millimeter wavelengths.\citep{jor07} This can be explained if the emission region is larger
than at optical wavelengths, encompassing more turbulent cells. Finally, the time-scale
of variability is generally shorter at higher frequencies between the radio and optical
bands.\citep{mj10} This can result from averaging of the fluctuations over more cells at
longer wavelengths.

\begin{figure}
\includegraphics[width=85mm]{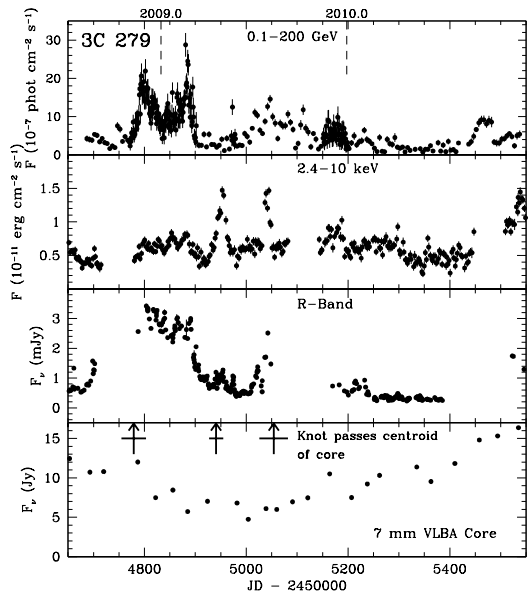}
\caption{Multiwavelength light curves of the quasar 
3C~279 from August 2008, when regular $\gamma$-ray monitoring by
{\it Fermi} began, to the end of 2010. Note the overall similarity of the $\gamma$-ray
and optical light curves but also the differences in the detailed variations. The vertical
arrows in the bottom panel indicate times when new superluminal radio knots passed the
core on 7 mm VLBA images. The flux of the core plotted includes the blended flux of
the new knot until it separates from the core by more than 0.1 milliarcsec. Data are
from Marscher et al. (in preparation).
}
\label{f1}
\end{figure}

Close examination of multi-waveband outbursts provides further evidence for chaotic behavior 
superposed on more systematic trends. The quasar 3C~279 is a prime example.
Figure \ref{f1} presents 2.4 years of light curves at $\gamma$-ray, X-ray, and optical
wavelengths, and in the 7 mm core (as seen in VLBA images).
Times when a new superluminal knot passes the 7 mm core coincide with various events:
(1) the beginning of the largest $\gamma$-ray and optical outburst, (2) an ``orphan'' X-ray
flare, and (3) a flare that occurred simultaneously at $\gamma$-ray, optical, and X-ray
energies. In addition, a major outburst in the core started before the $\gamma$-ray flare
at RJD 5460-5480 (where RJD = JD$-$2450000). Furthermore, a period of quiescence occurred
contemporaneously at
$\gamma$-ray, X-ray, and optical bands between RJD 5300 and 5440. It is clear that the
relationships among the wavebands are complex. Other blazars with multiple
outbursts---e.g.,PKS~1510$-$089 in 2009 \citep{mar10a,mar10b} and 3C~454.3 in late 2010 (Wehrle
et al., in preparation) exhibit similar complexities alongside obvious (and statistically significant) correlations. 

\begin{figure}
\includegraphics[width=85mm]{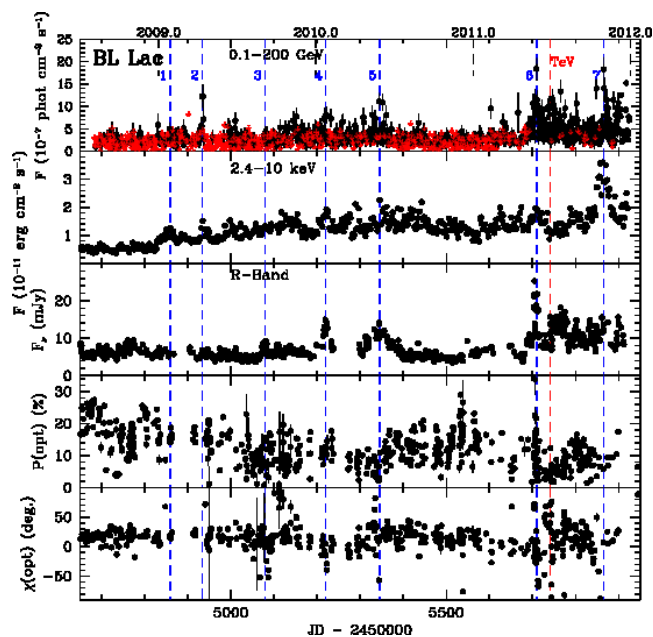}
\caption{Multi-waveband flux and degree $P$ and electric-vector
position angle $\chi$ of optical polarization vs. time for
BL Lac in 2011.  
Note rapid variations in $P$ and $\chi$, indicative of a turbulent magnetic field.
Data are from Marscher et al. (in preparation).
}
\label{f2}
\end{figure}

The difference between $\gamma$-ray and optical fluctuations is particularly noteworthy.
For magnetic field strengths between $\sim0.2$ and 1 G, as is usually inferred in
the optical synchrotron-emitting regions, electrons of the same energies (thousands times
the rest-mass energy) radiate at both optical wavelengths and 0.1-10 GeV energies. The
implication is that something besides the number of such electrons is changing to produce
the variations in flux. The synchrotron flux depends on the magnetic field strength $B$ as
well as its direction, number of radiating electrons $N_{\rm re}$, and Doppler factor
$\delta$, while the inverse Compton flux depends on only the latter two. If the seed
photons that are scattered arise from outside the jet, they are blueshifted in the plasma 
frame, in which case the $\gamma$-ray emission is more sensitive to $\delta$. Since the
$\gamma$-ray flux usually (but not always) fluctuates
with higher amplitude than does the optical flux, either $B$ is anti-correlated with
$N_0$ (which seems unlikely on physical grounds), the highest-energy electrons are
created mainly in cells with magnetic fields pointing almost toward the observer
(after correction for relativistic aberration), or the Doppler factor varies
considerably across the emission region. The last of these possibilities is a feature
of ``jet-in-jet'' models involving magnetic reconnection events \citep{gian09} or
relativistically turbulent motions.\citep{np12} These have been proposed to explain
extremely short time-scales of variability of the TeV $\gamma$-ray emission of some blazars.

BL Lac (Fig. \ref{f2}) provides another example of an object whose multi-waveband light
curves, polarization, and changes in sub-milliarcsecond-scale structure support the
scenario in which turbulence plays a major role in the nonthermal emission.
A number of $\gamma$-ray flares are seen, but their relationship to
optical and X-ray variations is complex. In 2011, the blazar became paticularly active
at all wavebands  starting about RJD = 5700,
as a new superluminal knot was passing through the core at 43 GHz (as occurred during
all of the main $\gamma$-ray flares seen in Fig. \ref{f2}). The optical polarization
became highly chaotic until just after a rapid TeV flare was observed by VERITAS.\citep{Ong11}
Another $\gamma$-ray flare near RJD = 5865 occurred during a major X-ray and radio
outburst after another superluminal knot appeared, while the
optical flux was fluctuating rapidly and $\chi$(opt) was again chaotic. Turbulence is
a natural way for the jet to produce these types of variations.\citep{mgt92}

\begin{figure}
\includegraphics[width=85mm]{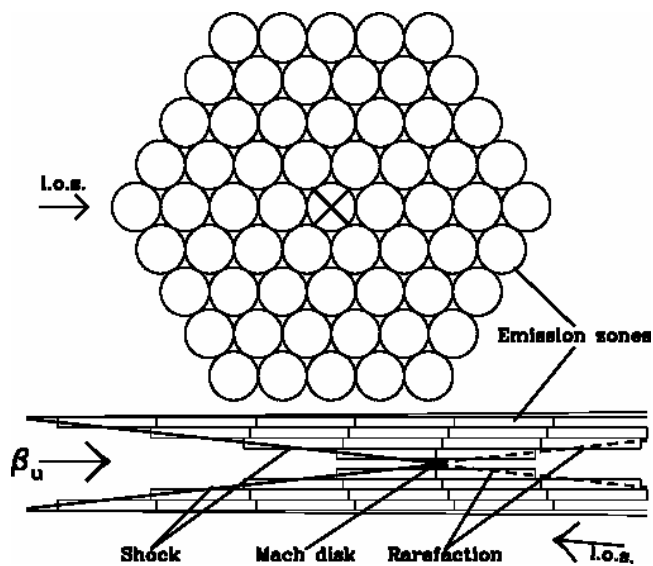}
\caption{Sketch of geometry adopted in the TEMZ model. {\it Top:} View down
the jet, whose cross-section is divided into many cylindrical cells. Actual
calculations displayed in Figs. 4-6 use 271 cells across the jet.
{\it Bottom:} Side-on view. Cylindrical cells
appear as rectangles. The conical shock compresses the flow \& accelerates electrons.
The Mach disk, if present, is at the axis, oriented transverse to the flow.
}
\label{f3}
\end{figure}

\section{TURBULENT EXTREME MULTI-ZONE (TEMZ) MODEL}

The author is developing a model in which much of the optical and high-energy radiation in a blazar is emitted parsecs from the central engine. The main physical features 
are a turbulent ambient jet plasma that passes through a conical standing shock wave (Òrecollimation 
shockÓ) in the jet. The model can generate short time-scales of optical and gamma-ray variability by 
restricting the highest-energy electrons radiating at these frequencies to a small fraction of the 
turbulent cells, perhaps those with a particular orientation of the magnetic field relative to the
shock front (Summerlin \& Baring 2012). Because of this, as well as radiative energy losses as the 
plasma advects beyond the shock front, the volume filling factor at high frequencies is relatively
low. Such a model is
consistent with the (1) red-noise power spectra of flux variations in blazars, (2) shorter
time-scales of variability of flux and polarization at higher frequencies, (3) mean polarization
levels as well as fractional deviations from the mean that are higher at optical than at lower frequencies, (4) apparent rotations in polarization position angle that are really just random
walks of the projected magnetic field direction, (5) breaks in the synchrotron spectrum by more
than the radiative loss value of 0.5, and (6) flares that are often sharply peaked or contain
multiple maxima, both of which are not reproduced by single- or few-zoned models. The
dependence of items 2-4 on frequency is directly related to the change in spectral index beyond
the break, according to the model.

\begin{figure}
\includegraphics[width=85mm]{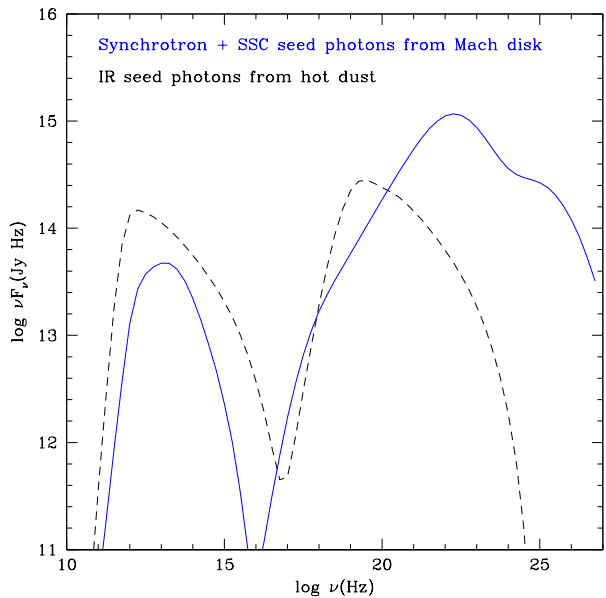}
\caption{Two simulated spectral energy distributions (SEDs) from the TEMZ code. In both cases,
the left side of the SED is from synchrotron radiation, while the right component
is from inverse Compton scattering. For the latter, the
seed photons are from IR from dust 
(dashed curve) and synchrotron plus SSC emission from the Mach disk (solid curve). 
}
\label{f4}
\end{figure}

The model includes synchrotron radiation, inverse Compton scattering (IC) of seed photons from hot
dust \citep[observed in 4C21.35][]{mm11}, and IC of synchrotron plus synchrotron self-Compton (SSC)
radiation from relatively slowly moving plasma in a Mach disk (shock oriented perpendicular to the axis
in cylindrically symmetric jets) at the conical shock's vertex.
If the Mach disk is present, this is the
dominant---and variable--- source of seed photons, since
the emission is Doppler boosted in the frame of the turbulent cells. The combined effects of
non-uniform electron energy distribution, different magnetic field orientations for different
turbulent cells, and light-travel delays often cause time lags and/or single waveband (``orphan'') 
flares (see Figs. 4-6 for sample output from preliminary computations). Thus, the model can
qualitatively reproduce some of the puzzling features of observed multi-waveband light curves
of blazars. In addition,
the discrete $\gamma$-ray/optical correlation function is similar to that observed in blazars,
with zero mean lag but considerable spread about this value.

\begin{figure}
\includegraphics[width=85mm]{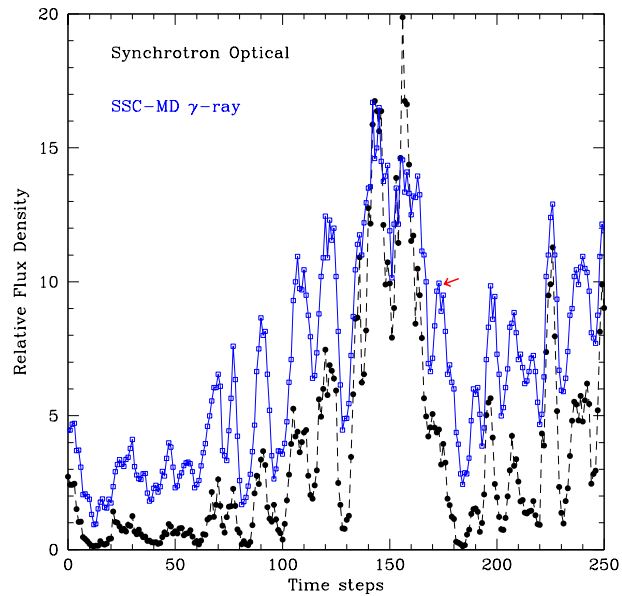}
\caption{Sample TEMZ light curves: optical synchrotron radiation (black, dashed curve) and
inverse Compton $\gamma$-ray flux. Seed photons for the latter are synchrotron
and first-order SSC emission from the Mach disk.
Red arrow points to an orphan $\gamma$-ray flare.
}
\label{f5}
\end{figure}

In the future, the author plans to develop the TEMZ model by adding the following capabilities:
\vskip 0mm\noindent
1. ``Peculiar'' velocities of the cells relative to the mean flow, as expected for turbulence
 or magnetic reconnections. This is probably necessary to
reproduce intra-day $\gamma$-ray variability with high apparent luminosities.\citep{np12,gian09}
\vskip 0mm\noindent
2. Full SSC calculation, with seed photon field for each cell including retarded-time
emission from all other cells. This is relevant mainly to BL Lac objects
with $\gamma$-ray to synchrotron integrated flux ratios $\lesssim 1$. It will require translation
of the code from Fortran to C++ and parallelization to run on a supercomputer.
\vskip 0mm\noindent
3. Refine the calculation of pair-production absorption at high $\gamma$-ray energies and
synchrotron self-absorption at low frequencies. The code currently does this crudely in order
to save computation time. These sources of opacity affect the spectral energy 
distribution (SED) at millimeter wavelengths and
at $\gamma$-ray energies above a few GeV.
\vskip 0mm\noindent
4.  Split each cell into $\sim 10$ sub-cells for a more refined calculation, to allow
intra-day variability. The current scheme assigns uniform
physical properties across each cell. Because of the acute angle of the conical shock,
this requires long, thin cells (see Fig. 3).
\vskip 0mm\noindent
5. Calculate the polarization of X-rays from inverse Compton scattering, in the anticipation
that this can be measured by GEMS (launch scheduled for 2014) for some blazars during
$\gamma$-ray/X-ray flares to add additional constraints on models.

\begin{figure}
\includegraphics[width=85mm]{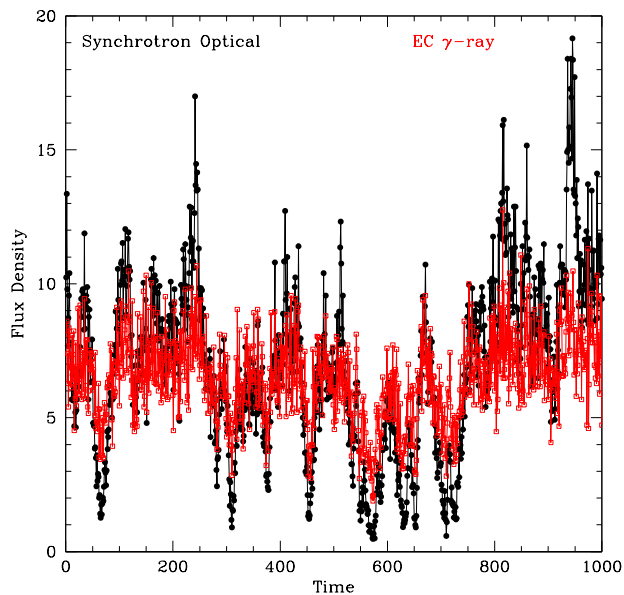}
\caption{Sample TEMZ light curves: optical synchrotron radiation (black) and inverse Compton
$\gamma$-ray emission (red). Seed photons for the latter are from thermal infrared emission
by a dusty torus.
}
\label{f6}
\end{figure}

\section{CONCLUSIONS}

While analyzing light curves as a series of distinct flares can be a convenient
way to study the physical conditions that exist at different flux states of
a blazar, the statistics of light curves indicates that noise processes dominate.
The degree and variability of linear polarization also lead directly to a picture
in which the magnetic field has a major chaotic component, which can result from
turbulence. The higher polarization and stronger variability at shorter
wavelengths further suggests that the volume filling factor of the emission decreases
with frequency. While this can arise from radiative losses after electrons
are energized at a shock front \citep{mg85,mgt92}, the break in the synchrotron
spectrum is usually greater than the change from spectral index $\alpha$ to
{($\alpha+0.5$)} expected from this process. A sharper break can occur if electrons
can be accelerated to higher energies only over a smaller fraction of the total
volume.

The author is incorporating these requirements into a qualitatively new paradigm
for both the multi-waveband variability of blazars and the appearance of
superluminal radio knots. Stochastic variations in the jet flow --- changes
with time of the magnetic field and electron density at the jet input, perhaps
driven by fluctuations in the accretion flow or magnetic field configuration
in the central engine --- combined with turbulence within the flow
cause the variability in emission as the plasma crosses standing shocks on
parsec scales in the jet. Seed photons for inverse Compton scattering can come
from thermal infrared emission by a hot dust torus, synchrotron radiation from
the turbulent plasma, or synchrotron plus first-order SSC radiation emitted
beyond a Mach disk, which is a small, strong shock oriented perpendicular to the
jet axis that slows the flow down to subsonic speeds. The emission from the Mach
disk is highly blueshifted in the frame of the plasma that crosses the conical
shock that extends across most of the jet cross-section.

The rapid variability in the TEMZ model occurs because of the small volume
filling factor of the highest-frequency emission, as well as the fluctuations
in magnetic field, electron density, and maximum electron energy from one
turbulent cell to another. The addition of velocity structure in the turbulence
will accentuate the variability on the shortest time-scales. After developing the
numerical TEMZ code further, the author will run it over a wide range of
parameter space in order to compare the simulated light curves and dynamic SEDs
with those that are observed in blazars. 
\bigskip
\begin{acknowledgments}
This research is supported in part by NASA through Fermi grants NNX08AV65G, NNX10AO59G, and NNX11AQ03G, and by NSF grant AST-0907893. The VLBA is an instrument of the National Radio Astronomy Observatory,
a facility of the National Science Foundation operated under cooperative agreement by Associated Universities, Inc.

\end{acknowledgments}

\bigskip 

\end{document}